\documentclass[aps,prb,reprint,groupedaddress,amsfonts,amssymb,amsmath]{revtex4-2}

\usepackage{graphicx}
\usepackage{dcolumn}
\usepackage{bm}
\usepackage[mathlines]{lineno}
\usepackage{siunitx}
\usepackage{booktabs}
\usepackage{color}
\usepackage{multirow}
\usepackage{mathtools}

\begin{document}

\title{Topological nodal line in ZrTe$_2$ demonstrated by nuclear magnetic resonance}

\author{Yefan Tian}
\affiliation{Department of Physics and Astronomy, Texas A\&M University, College Station, TX 77843, USA}
\author{Nader Ghassemi}
\affiliation{Department of Physics and Astronomy, Texas A\&M University, College Station, TX 77843, USA}
\author{Joseph H. Ross, Jr.}
\email{jhross@tamu.edu}
\affiliation{Department of Physics and Astronomy, Texas A\&M University, College Station, TX 77843, USA}

\date{\today}

\begin{abstract}
In this work, we report nuclear magnetic resonance (NMR) combined with density functional theory (DFT) studies of the transition metal dichalcogenide ZrTe$_2$. The measured NMR shift anisotropy reveals a quasi-2D behavior connected to a topological nodal line close to the Fermi level. With the magnetic field perpendicular to the ZrTe$_2$ layers, the measured shift can be well-fitted by a combination of enhanced diamagnetism and spin shift due to high mobility Dirac electrons. The spin-lattice relaxation rates with external field both parallel and perpendicular to the layers at low temperatures match the expected behavior associated with extended orbital hyperfine interaction due to quasi-2D Dirac carriers. In addition, calculated band structures also show clear evidence for the existence of nodal line in ZrTe$_2$ between $\Gamma$ and A. For intermediate temperatures, there is a sharp reduction in spin-lattice relaxation rate which can be explained as due to a reduced lifetime for these carriers, which matches the reported large change in mobility in the same temperature range. Above 200 K, the local orbital contribution starts to dominate in an orbital relaxation mechanism revealing the mixture of atomic functions. 
\end{abstract}

\maketitle

\section{Introduction}

In recent years, there has been great interest in layered transition metal dichalcogenides (TMDCs), comprised of a wide range of transition metal (Mo, W, Ta, Zr, Hf, \textit{etc.}) and chalcogen (S, Se, or Te) elements. The TMDC family offers platforms for exploring striking physical phenomena and exotic electronic device applications \cite{manzeli20172d}. Among TMDCs, ZrTe$_2$ has been relatively little investigated; however, recent work \cite{machado2017evidence,tsipas2018massless,kar2020metal} has indicated interesting topological features in this material both in the normal state and as a doped superconductor. Also, other zirconium tellurides have been of considerable interest. For instance, ZrTe$_5$ shows interesting topological properties and unique physical properties such as chiral magnetic effect \cite{li2016chiral} and three-dimensional quantum Hall effect \cite{tang2019three}. ZrTe$_5$ also exhibits a topological phase transition separating the strong and weak topological insulator states \cite{manzoni2016evidence,fan2017transition,tian2019dirac} with a temperature-driven valence and conduction band inversion associated with this topological phase transition \cite{manzoni2016evidence}. The layered material ZrTe$_3$ has also been long studied due to interesting behavior such as a charge density wave phase transition \cite{takahashi1984transport}. Recently, theoretical calculations indicate distinctive topological behavior in ZrTe, which possesses triple-point fermions coming from the three-fold degenerate crossing points formed by the crossing of a double-degeneracy band and a nondegeneracy band \cite{zhu2016triple}.

\textcolor{black}{Regarding ZrTe$_2$, theoretical predictions from several groups give rather different results \cite{reshak2004theoretical,guo2014tuning,machado2017evidence,tsipas2018massless,muhammad2019transition,kar2020metal}, including several \cite{reshak2004theoretical,machado2017evidence} predicting ZrTe$_2$ to be a simple metal; however, recent experimental evidence appears to contradict this result. In addition, recent ARPES studies \cite{kar2020metal} have presented evidence of massless Dirac fermions observed in the ZrTe$_2$ bulk phase, while recent DFT calculations \cite{tsipas2018massless,kar2020metal} have also supported the topological semimetal prediction, thus indicating that ZrTe$_2$ may have promising prospects for quantum device applications. NMR spectroscopy is particularly sensitive to electronic carriers near the Fermi level, based on the observation of spectral shifts and also nuclear relaxation times, and thus provides an effective means to characterize the behavior of the Dirac carriers in this system.}

\begin{figure}
\includegraphics[width=0.9\columnwidth]{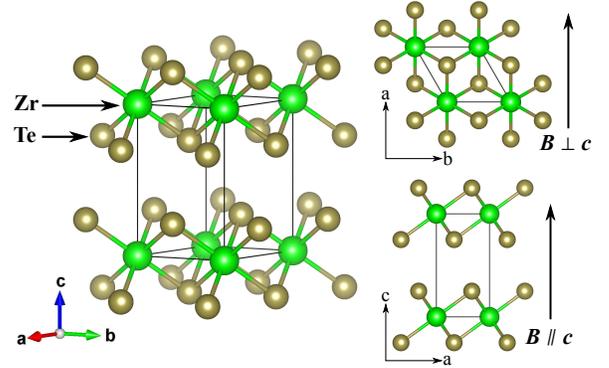}%
\caption{\label{structure} Crystal structure of 1T-ZrTe$_2$ with P-3m1 space group, showing van der Waals-bonded layered structure.}
\end{figure}

\textcolor{black}{In this work, we have studied ZrTe$_2$ using NMR techniques combined with DFT computations, and the results indicate a strongly diamagnetic response of Dirac carriers circulating within the ZrTe$_2$ layers, but with a quasi-2D behavior that becomes modified as the temperature increases at low $T$. By observing the differences in NMR shifts and spin-lattice relaxation rates for both the $B\parallel c$ and $B\perp c$ orientations, we find that the low-temperature results correspond to a nodal line extending in the direction perpendicular to the layers.}

\begin{figure*}
\includegraphics[width=1.6\columnwidth]{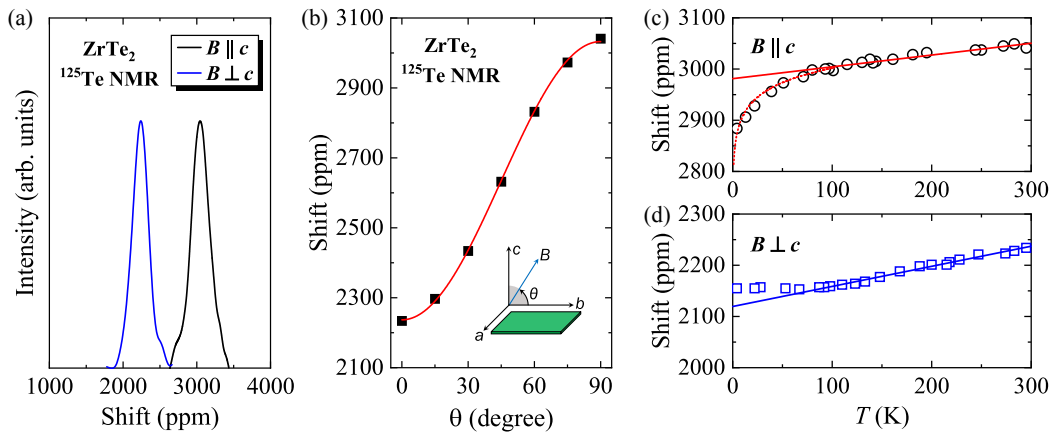}%
\caption{\label{shift} (a) $^{125}$Te lineshapes of ZrTe$_2$ at room temperature. (b) Angular dependence of shift at room temperature. The red solid curve is a fit to Eq.~(\ref{eq_angle}). Shift vs temperature for (c) $B\parallel c$ (magnetic field perpendicular to the layers) with linear and $\ln(T)$ curves as guides to the eye and (d) $B\perp c$ (magnetic field parallel to the layers).}
\end{figure*}

\section{Experimental and computational methods}

The ZrTe$_2$ single crystals (crystal structure shown in Fig.~\ref{structure}) were prepared using chemical vapor transport. The stoichiometric mixture of Zr and Te powder was sealed in a quartz tube with iodine being used as transport agent (2 mg/cm$^3$). Plate-like single crystals with metallic luster were obtained via vapor transport growth with a temperature gradient from 950\,$^\circ$C to 850\,$^\circ$C. Cameca SXFive microprobe measurements indicate a uniform phase Zr$_{0.99}$Te$_2$.

NMR experiments utilized a custom-built spectrometer at a fixed field $B\approx9$ T. Many individual crystals were stacked with the $c$ axes aligned and the sample was measured with the field parallel to $c$ ($B\parallel c$) and in the basal plane ($B\perp c$). The $a$ axis orientation was not identified for these crystals. $^{125}$Te \textcolor{black}{(nuclear spin $I=1/2$ and gyromagnetic ratio $\gamma=-8.51\times10^7$ rad\,s$^{-1}$\,T$^{-1}$)} shifts were calibrated by aqueous Te(OH)$_6$ and adjusted for its $\delta=707$ ppm paramagnetic shift to the dimethyltelluride standard \cite{inamo1996125te}.

The band structure and density of states calculations were carried out in the framework of the density functional theory (DFT) by employing the APW plus local orbital (APW+lo) method \cite{madsen2001efficient} with the PBE potential \cite{perdew1996generalized} as implemented in the \textsc{WIEN2k} code \cite{blaha2020wien2k}. A mesh of 1000 $k$-points was employed in the irreducible wedge of the hexagonal Brillouin zone [see Fig.~\ref{dft}(d)] corresponding to the grids of $10 \times 10 \times 10$ in the Monkhorst-Pack \cite{monkhorst1976special} scheme. The cutoff parameter of $k_\mathrm{max} = 7/R_\mathrm{MT}$ inside the interstitial region was used for the expansions of the wave functions in terms of the plane waves.

\section{Experimental and computational results}

\subsection{Shift}

Consistent with the single local environment for Te in the 1T-ZrTe$_2$ structure, there is only one peak observed in the $^{125}$Te spectra as shown in Fig.~\ref{shift}(a). The angular dependence of the NMR shift (with $\theta$ defined between the $ab$ layer and the magnetic field $B$) is shown in Fig.~\ref{shift}(b). The room-temperature shift was fitted [red curve in Fig.~\ref{shift}(b)] to 
\begin{equation} \label{eq_angle}
K=K_\mathrm{iso}+\frac{3\cos^2\theta-1}{2}\cdot\Delta K,
\end{equation}
where $K_\mathrm{iso}=2767\pm3$ ppm is the isotropic shift and $\Delta K=-530\pm4$ ppm. By symmetry, the shift will not depend on orientation in the basal plane, which is confirmed by the absence of additional inhomogeneous line broadening for this orientation [Fig.~\ref{shift}(a)]. Ref.~\cite{orion1997solid} gives $\delta_\mathrm{iso}=1825$ ppm with Te(OH)$_6$ as reference, which corresponds to 2532 ppm, a similar shift as reported here, considering the large width measured in Ref.~\cite{orion1997solid}.

Figs.~\ref{shift}(c) and \ref{shift}(d) show the temperature dependence of the $^{125}$Te shift for $B\parallel c$ and $B\perp c$ ($K_{\parallel c}$ and $K_{\perp c}$), respectively. The shifts were obtained by identifying the highest intensity position of the measured single-peak $^{125}$Te spectra. \textcolor{black}{Both $K_{\parallel c}$ and $K_{\perp c}$ decrease monotonically vs $T$, with $K_\mathrm{iso}$ corresponding to the linear fits [shown in Figs.~\ref{shift}(c) and \ref{shift}(d)] changing by 0.34 ppm/K. At low $T$, $K_{\parallel c}$ shows a sharp decrease as $T$ approaches zero, while for $K_{\perp c}$, there is a clear change in the opposite direction close to 50 K, where the shift is nearly temperature independent. These results are indicative of quasi-2D Dirac-node behavior as is discussed in Sec.~\ref{analysis_shift}.} 

The carrier concentration shown in Ref.~\cite{kar2020metal} is in the order of $10^{19}$ cm$^{-3}$, which presents the fact that the large measured shifts are mostly chemical shifts due to electronic states away from the Fermi energy ($\varepsilon_F$); however, the temperature-dependence is dominated by Knight shifts due to carriers at $\varepsilon_F$, and for convenience we label observed shift, which is the sum of these shift terms, as $K$.

\subsection{Spin-lattice relaxation}

\begin{figure}
\includegraphics[width=0.9\columnwidth]{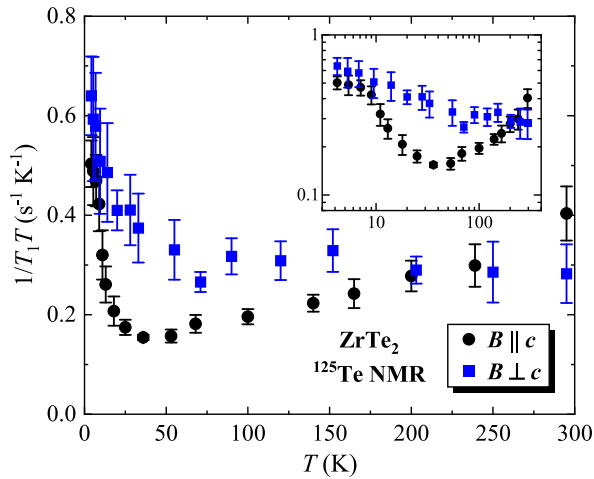}%
\caption{\label{T1TvsT} $1/T_1T$ vs $T$ for both orientations $B\parallel c$ (perpendicular to the layers) and $B\perp c$ (parallel to the layers). Inset: $1/T_1T$ vs $T$ in log scale.}
\end{figure}

Spin-lattice relaxation results, measured by inversion recovery, could be well fitted to a single exponential $M(t)=(1-Ce^{-t/T_1})M(\infty)$, giving $1/T_1T$ values shown in Figs.~\ref{T1TvsT}(a) and \ref{T1TvsT}(b). The results decrease rapidly at low temperatures as $T$ increases, especially $(1/T_1T)_{\parallel c}$, \textcolor{black}{which changes rather quickly at temperatures near 15 K.} Near 50 K, which is also the temperature at which $K_{\perp c}$ exhibits a change in behavior, the relaxation results also exhibit a characteristic change, with $1/T_1T$ leveling off, and $1/T_1T$ exhibiting a minimum near 40 K and then steadily increasing. In metals, $1/T_1T$ is often dominated by $s$-electron Fermi contact and proportional to $g^2(\varepsilon_F)$. \textcolor{black}{However, similarly to ZrTe$_5$ \cite{weng2014transition} we find that the Dirac states in ZrTe$_2$ are dominated by Te $p$-orbitals, along with Zr $d$-states, as confirmed by the DFT results which are described in the next section. These produce a dominant orbital contribution to the $1/T_1T$, and we will further demonstrate that the largest term is due to the high-mobility Dirac carriers.}

\subsection{DFT computations} \label{sec_dft}

\begin{figure*}
\includegraphics[width=1.9\columnwidth]{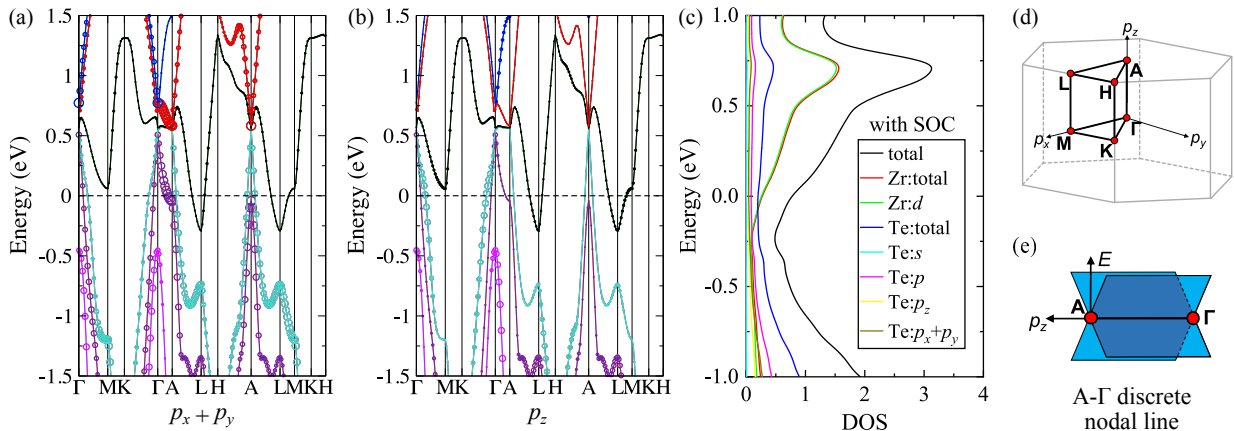}%
\caption{\label{dft} Band structures of ZrTe$_2$ with spin-orbit coupling, with superposed circles showing weights for (a) $p_x+p_y$ and (b) $p_z$ Te orbitals. \textcolor{black}{The dashed lines represent the Fermi level.} The circle size represents the partial state density of Te. (c) Density of states for ZrTe$_2$. (d) 3D view of the hexagonal Brillouin zone with high-symmetry points. (e) Sketch of discrete nodal line between $\Gamma$ and A.}
\end{figure*} 

From reports by several groups \cite{reshak2004theoretical,guo2014tuning,machado2017evidence,tsipas2018massless,muhammad2019transition,kar2020metal}, there have been some conflicts about the topological nature of ZrTe$_2$ as detected in DFT results. Ref.~\cite{muhammad2019transition} suggests a semimetallic state of ZrTe$_2$ without any topological nature. Ref.~\cite{kar2020metal} suggests ZrTe$_2$ is a topological semimetal, consistent with its ARPES results. Both Refs.~\cite{tsipas2018massless,kar2020metal} indicate a Dirac point at $\Gamma$ with the Dirac node close to the chemical potential and an electron pocket at M in the conduction band. The lattice parameters used in Ref.~\cite{kar2020metal} are about 1-2\% expanded from experimental values. However, these parameters were obtained from a DFT energy optimization, and they provided an approximate match for the reported ARPES results, with the calculated Dirac node roughly 0.5 eV higher in energy than \textcolor{black}{what is actually observed by ARPES, and with larger calculated overlaps of the pockets at L and M} than what is observed. Ref.~\cite{muhammad2019transition} included a correction for the van der Waals interaction, leading to a much smaller overlap at the L and M points; however, a large gap opened throughout the Brillouin zone, in seeming contradiction with magnetotransport results \cite{wang2019magnetotransport} as well as APRES results \cite{kar2020metal}. It is likely that the well-known difficulty in predicting band energies near the gap in standard GGA functionals such as PBE is responsible for the discrepancies between the calculated results and the observation. \textcolor{black}{In TMDCs specifically DFT is well-known to underestimate the band gaps \cite{cheiwchanchamnangij2012quasiparticle,pike2018vibrational}.} For further investigation we used the lattice parameters of Ref.~\cite{kar2020metal} ($a=3.909$ \si{\angstrom} and $c=6.749$ \si{\angstrom}) for DFT calculations, with the understanding that \textcolor{black}{the $\varepsilon_F$ position is much closer to the Dirac node than predicted.} 

Results of the DFT calculations, with spin-orbit coupling included, are shown in Figs.~\ref{dft}(a)-(c). The nearly-dispersionless band from $\Gamma$ to A connects to Dirac-like features at $\Gamma$ (as previously identified \cite{tsipas2018massless,kar2020metal}) and also at A, and this band is doubly degenerate except for a gap of about 20 meV very close to $\Gamma$, identified \cite{kar2020metal} as associated with a band inversion. The mapping in reciprocal space, and a schematic of the nodal line between $\Gamma$ and A, are demonstrated in Figs.~\ref{dft}(d) and \ref{dft}(e). Also note that the partial DOS results show that Te $p$-orbitals mostly locate at these Dirac bands away from the node while Zr $d$-orbitals dominate at the node itself, and the Zr orbitals dominate the electron pockets at L and M. There is also a separate high-dispersion band crossing $\Gamma$ just below the node energy.

As an estimate of the Fermi velocity for the Dirac nodal line, we analyzed the linear slope in the $\Gamma$-M and A-L directions leading up to the nodal line according to $\varepsilon=\hbar v_F k$, and obtained 6.9 and $6.5\times 10^5$ m/s. Based on these values, \textcolor{black}{which are typical for Dirac semimetals \cite{armitage2018weyl}, we will use the mean value, $6.7\times 10^5$ m/s, for further analysis of the Dirac-carrier behavior.} A similar value was estimated for the monolayer case \cite{tsipas2018massless}. The extra pockets at L and M contain ordinary electrons, and the existence of both Dirac and ordinary electrons at $\varepsilon_F$ leads to additional complexity in this case, although experimental indications \cite{tsipas2018massless,kar2020metal} point to a much smaller overlap between the M pocket and the Dirac valence band than what is calculated. With the $\sqrt{\varepsilon}$ type density of states near $\varepsilon_F$ dominated by the M pocket we fitted to $g(\varepsilon)=\sqrt{(2\varepsilon(m^*)^3)}/(\pi^2\hbar^3)$ and obtained an estimate of $m^*=1.7\,m_e$ for this pocket. In the model discussed in Sec.~\ref{analysis_shift}, the position of Fermi level is near the edge of this pocket, and very close to the nodal line.

\section{Discussion and analysis} \label{analysis}

\subsection{Knight shift} \label{analysis_shift}

As shown in Figs.~\ref{shift}(c) and \ref{shift}(d), there is an obvious difference between the measured shifts of $B\parallel c$ and $B\perp c$ orientations, especially at low temperatures. The observed low-$T$ divergence for $K_{\parallel c}$ follows approximately a $\ln(T)$ curve, characteristic of the divergent orbital susceptibility for Dirac semimetals \cite{okvatovity2019nuclear,maebashi2019nuclear}, although the absence of the corresponding behavior for $K_{\perp c}$ points to a quasi-2D Dirac semimetal rather than 3D point-node behavior. 

\begin{figure*}
\includegraphics[width=2\columnwidth]{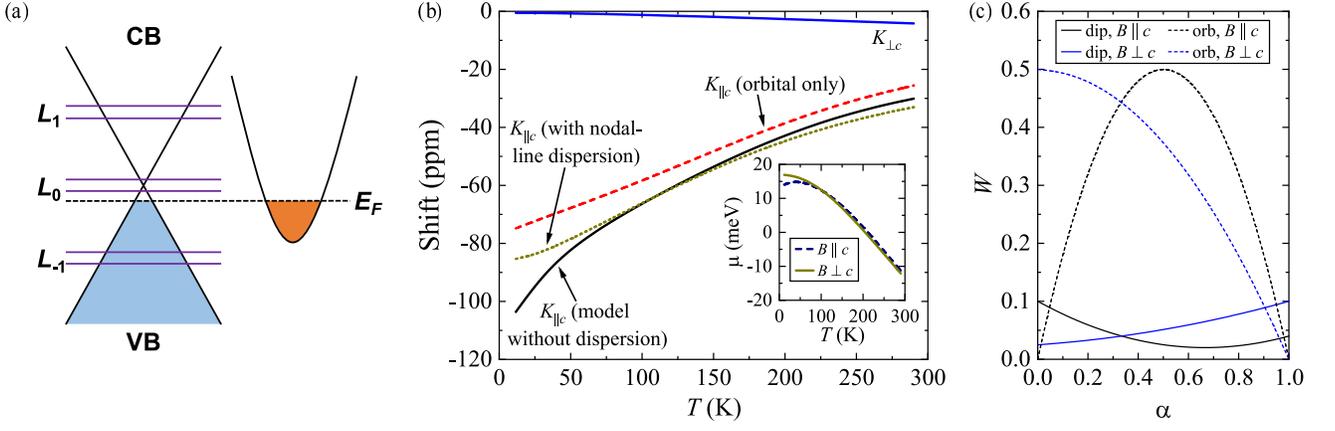}%
\caption{\label{discussion} (a) Sketch of Dirac band and electron pocket. (b) Simulated shifts for both orientations. Inset: chemical potential vs $T$. (c) $W=1/2T_1$ is the dipolar and orbital relaxation rates divided by $2\pi(\gamma_e\gamma_n\hbar^{3/2})^2g^2(\varepsilon_F)k_BT\langle r^{-3}\rangle^2$. $\alpha$ is the mixture of orbitals ($p_x + p_y$) vs $p_z$.}
\end{figure*}

To analyze this situation, first we note that the shifts will be largely due to the dominant $p$-electrons for Te in ZrTe$_2$, contributing a combination of core polarization and spin-dipolar shifts, which are due to electron spin mechanisms, as well as orbital shifts, with the latter likely dominated by the large bulk orbital response of the Dirac electrons rather than due to local orbitals. The core polarization mechanism normally contributes an isotropic shift (the same sign for both orientations) and the spin-dipolar, anisotropic shift [second term in Eq.~(\ref{eq_angle})]. However, the absence of divergent behavior for $B\perp c$ points to a different physical mechanism for the two orientations rather than shift anisotropy, and thus, we analyze the $B\parallel c$ divergence in terms of the spin response of quasi-2D Dirac electrons due to the separation of Landau levels with $B\parallel c$, plus an orbital shift dominated by quasi-2D orbital currents confined to the basal plane.

For quantitative comparison, first we consider the case of a 3D point node. The Knight shift due to the orbital interaction in a 3D massless Dirac electron case can be expressed as \cite{maebashi2019nuclear},
\begin{equation} \label{eq_shift}
K=K_0-\bigg[\frac{\mu_0v_F e^2}{6\pi^2\hbar}\ln\bigg(\frac{W}{\mathrm{max}\{k_BT,|\mu|\}}\bigg)\bigg]\textcolor{black}{(1-N_D)},
\end{equation}
where $K_0$ is a $T$-independent term, \textcolor{black}{$\mu$ is the chemical potential measured from the Dirac node}, $W$ is a bandwidth cutoff and \textcolor{black}{$N_D$ is demagnetizing factor. $N_D$ can be significant for the orbital hyperfine contribution of extended Dirac carriers, and in fact in the pure 2D limit the shift due to this mechanism will vanish \cite{dora2009unusual}.} Note this is the low field case. For $v_F$, we used $v_F=6.7\times10^5$ m/s from the DFT results (Sec.~\ref{sec_dft}). Considering the demagnetizing effect, the overall sample size (around $2\times2\times0.5$ mm$^3$) implies a demagnetizing factor of approximately \textcolor{black}{$N_D=0.8$} for such a bulk-susceptibility contribution for the $B\parallel c$ orientation. \textcolor{black}{Using these values, and assuming that $k_BT$ dominates in the logarithm of Eq.~(\ref{eq_shift}), we obtain a difference in shift of less than 1 ppm between the temperatures 10 K and 100 K, much less than what is observed. Or, if changes in $\mu$ are on the order of $k_BT$, the results will be similarly small.}

\subsection{\textcolor{black}{Quasi-2D model for Knight shift}} \label{model_shift}

As alternative we consider the shift due to the diamagnetic currents of a Dirac nodal line oriented along the $c$ direction. In this quasi-2D case, currents are confined to the basal plane, and the diamagnetic response is equivalent to that of a 2D Dirac gas, for which we follow the treatment used for graphene \cite{li2015field}. Also note that the effect vanishes for $B\perp c$, due to the absence of high mobility circulating currents perpendicular to the plane. For ZrTe$_2$ we modeled this system as including a quasi-2D Dirac line, with the addition of a normal electron pocket crossing the node energy ($\varepsilon_\mathrm{node}$), as indicated by DFT calculations and by ARPES measurements \cite{kar2020metal}.

\textcolor{black}{First, we calculate the chemical potential ($\mu$).} For the normal electron pocket we assumed an effective mass $m^*/m_e=1$, close to the estimate for the pocket at M in DFT calculations (\textcolor{black}{Sec.~\ref{sec_dft}). Also for the perpendicular Fermi velocity we used the result obtained from DFT}, $v_\perp=6.7\times10^5$ m/s, which in the 9 T NMR field perpendicular to the layers gives Landau-level energies $\varepsilon_\mathrm{LL}(N)=\pm\sqrt{(2e\hbar v_\perp^2 B|N|)}= \pm 73\sqrt{(|N|)}$ meV, and a volume density of carriers per spin level $n_\mathrm{LL} = B/(\Phi_0 c)=3.3\times10^{18}$ cm$^{-3}$, where $\Phi_0=4.14\times 10^{-15}$ T\,m$^2$ \textcolor{black}{is the magnetic flux quantum.} The gyromagnetic ratio is not known for these carriers, so we assumed $g=2$. Also we assumed that a fixed density of carriers $n_\mathrm{total} = 10^{19}$ cm$^{-3}$ estimated from ARPES results \cite{kar2020metal} is divided between these band features. To solve for the chemical potential we specified,
\begin{multline} \label{ntotal}
n_\mathrm{total}=\int_0^\infty f(\varepsilon,\mu)g_\mathrm{CB}(\varepsilon)d\varepsilon \\
+\sum_{s=-1/2}^{1/2}\sum_{N=\textcolor{black}{-\infty}}^{\textcolor{black}{\infty}}n_\mathrm{LL} f(\varepsilon_N, \mu)-n_\mathrm{LL}-\sum_{s=-1/2}^{1/2}\sum_{N=\textcolor{black}{-\infty}}^{-1}n_\mathrm{LL},
\end{multline}
where $g_\mathrm{CB}(\varepsilon)=\sqrt{(2\varepsilon {m^*}^3)}/(\pi^2\hbar^3)$ is the density of states in the normal-carrier pocket with its minimum set to $\varepsilon=0$, $\varepsilon_N=\varepsilon_\mathrm{node}+\mu_B gBs+\varepsilon_\mathrm{LL} (N)$ represents the Landau level energies, and $f(\varepsilon,\mu)=1/[1+e^{(\varepsilon-\mu)/k_BT}]$ is the Fermi function. \textcolor{black}{The extra term $n\mathrm{LL}$ comes about because the lower $N=0$ level is derived from the hole states, and we apply level quantization only to the Dirac states for which the large $v_F$ pushes these states into the quantum limit. In the finite sums, we chose a very large cutoff for which the sums are numerically well-converged.} In the $B\perp c$ case for which the Landau levels collapse, we replaced the sum over Landau levels in Eq.~(\ref{ntotal}) with an integral over the 2D Dirac density of states $g_D(\varepsilon)=|\varepsilon-\varepsilon_\mathrm{node}\pm \mu_B gBs|/[\pi c(\hbar v_\perp)^2]$ per spin, also normalized for hole states similarly to the last term in Eq.~(\ref{ntotal}). Solving for $\mu(T)$, we obtained the results shown in the inset of Fig.~\ref{discussion}(b), for the case $\varepsilon_\mathrm{node}=12$ meV. Because of the significant carrier density $n_\mathrm{LL}$ at each Landau level energy including $N = 0$, the $B\parallel c$ field tends to pull $\mu$ into $\varepsilon_\mathrm{node}$ at low temperature \cite{mikitik2020crossing}, as can be seen from the results shown in the inset of Fig.~\ref{discussion}(b). Recently anomalous magnetotransport effects were also identified in a layered Dirac material due to field-induced alignment of the chemical potential \cite{xu2020interlayer}.

\textcolor{black}{We next calculate the diamagnetic susceptibility, $\chi=\mu_0 \partial M/\partial B$, and its contribution to the NMR shift, $K=\chi(1-N_D)$. The magnetization for $B\parallel c$ is $M=-(1/V)\partial \Omega/\partial B$ \cite{li2015field}} with the grand potential volume density given by
\begin{equation} \label{Omega}
\Omega/V=-k_BTn_\mathrm{LL}\sum^m_{N=-m}\ln[1+e^{(\varepsilon_N-\mu)/k_BT}],
\end{equation}
with $m$ a numerical cutoff for the sum. For numerical calculation of the $B$ derivative, we adopted the method described in Ref.~\cite{li2015field} to normalize for the $B$-dependence caused by the numerical cutoff $m$. Using the $\mu(T)$ results \textcolor{black}{shown in the inset of Fig.~\ref{discussion}(b)}, we thus arrived at an estimation of $\chi$ for the $B\parallel c$ case. \textcolor{black}{For the $B\perp c$ case, the diamagnetic contribution is zero since there is no splitting into Landau levels.} Using the demagnetizing factor $\textcolor{black}{N_D}=0.8$ estimated for our sample for $B\parallel c$, we arrived at the bulk-diamagnetic contribution to $K_{\parallel c}$ shown by the dashed curve in the main plot of Fig.~\ref{discussion}(b). Note that in the $B$-derivative of $\Omega/V$ we included changes in $n_\mathrm{LL}$ and $\varepsilon_\mathrm{LL} (N)$, but not in the numerical solutions $\mu(T)$. The difference should be small, since for most of the temperature range the CB pocket determines the position of $\mu$, while at low temperatures the results have the linear-$T$ behavior equivalent to the case that $\mu$ is fixed at $\varepsilon_\mathrm{node}$ \cite{li2015field}, due to the pulling effect of the magnetic field. 

To calculate the spin contribution to the shift, we first calculated the Dirac-electron spin density as
\begin{equation} \label{nspin}
n_\mathrm{spin}=\sum_{s=-1/2}^{1/2}\textcolor{black}{2s}\sum_{N=\textcolor{black}{-\infty}}^{\textcolor{black}{\infty}}n_\mathrm{LL}f(\varepsilon_N, \mu),
\end{equation}
both for $B\parallel c$ and $B\perp c$ using the corresponding $\mu(T)$ values \textcolor{black}{shown in the inset of Fig.~\ref{discussion}(b)}. Assuming the core-polarization hyperfine contribution dominates for the Te $p$-electrons participating in the Dirac node, we used the estimated \cite{carter1977metallic} hyperfine field $B_\mathrm{cp}^\mathrm{HF}=-15$ T in calculating the spin shift as $K_\mathrm{spin}=n_\mathrm{spin} (B_\mathrm{cp}^\mathrm{HF}/9$ T$)(V_\mathrm{cell}/2)$, with 9 T the applied NMR field and the sample volume per Te atom given by $V_\mathrm{cell}/2 = 50$ \si{\angstrom}$^3$. The results were added to the calculated $T$-dependent diamagnetic orbital shift, giving the spin+orbital result plotted in Fig.~\ref{discussion}(b) (lowest curve). The results are comparable to the observed shift behavior and have the same general temperature dependence. Since there is considerable likelihood that $g$ differs from 2 \cite{jeon2014landau,fu2016observation,liu2016zeeman,hu2017nearly}, we did not attempt a quantitative fitting; however, it appears that this model correctly captures the low-$T$ behavior, and that a combination of spin susceptibility and orbital diamagnetism, both strongly enhanced in the quantum limit for the $B\parallel c$ orientation, are responsible for the observations.

\textcolor{black}{Comparing to the 3D case discussed earlier [Eq.~(\ref{eq_shift})], we can thus understand the enhanced effect for the quasi-2D case as due to two effects. First, the lack of Landau level dispersion in 2D means that the density of states is changed considerably by the field, which allows for a large spin polarization since a large number of states is concentrated at discrete energies. Secondly, this concentration of states in energy also enhances the diamagnetic response obtained from Eq.~(\ref{Omega}). Also note that the estimated $\mu(T)$ obtained from Eq.~(\ref{ntotal}) [Inset of Fig.~\ref{discussion}(b)], should be little changed in the 3D case because of the large role of $g_\mathrm{CB}(E)$, and indeed these changes in $\mu(T)$ are on order of $k_BT$, confirming the estimate in Sec.~\ref{analysis_shift} of the small expected shift in that case.}

Note that in the DFT results (Fig.~\ref{dft}), a small dispersion appears in the nodal line, with the changes covering a range of approximately 20 meV between $\Gamma$ and A. To model the effect of this behavior, we added a simple linear dispersion to the $\varepsilon_\mathrm{node}$ position. This was done by modifying the sum over Landau level numbers $N$ in Eqs.~(\ref{ntotal})-(\ref{nspin}), replacing the summands having fixed $\varepsilon_\mathrm{node}$ by an integrated square distribution covering a range $\varepsilon_\mathrm{node} \pm10$ meV, and repeating the numerical calculations described above with otherwise identical parameters. This yielded the spin+orbital shift result shown in the dotted curve in Fig.~\ref{discussion}(b): the main effect is a softening of the spin contribution as $T$ approaches zero; however, the calculated magnitude is similar to that of the completely dispersionless case.

\subsection{Relaxation mechanisms}

The low-$T$ $1/T_1$ results exhibit an anisotropy and temperature dependence which does not match the corresponding behavior of the measured shifts. Thus, we expect the $T_1$ behavior is not a result of a Korringa-type spin contribution \cite{slichter2013principles} which would be expected in that case. \textcolor{black}{However, in contrast to the spin contribution}, the orbital shift and $T_1$ are not governed by a Korringa relation \cite{okvatovity2019nuclear}, and the behavior in the low-$T$ limit matches what is predicted \cite{maebashi2019nuclear,lee1991relaxation} for the quasi-2D orbital case due to a mechanism governed by high-mobility carriers \textcolor{black}{which we denote here as the extended orbital mechanism, since carriers far from the nucleus dominate this process.} For the quasi-2D free-electron gas (i.e., metallic layers where the electrons behave as a 2D free-electron gas), Lee and Nagaosa obtained the relaxation rates due to this mechanism when the magnetic field is applied parallel and perpendicular to the layers \cite{lee1991relaxation}, which corresponds to a ratio between $(1/T_1T)_{\parallel c}$ and $(1/T_1T)_{\perp c}$ of $2:3$. As shown in Fig.~\ref{T1TvsT}, excluding a $T$-independent background, the low-$T$ $(1/T_1T)_{\parallel c}$ and $(1/T_1T)_{\perp c}$ reaches a ratio close to $2:3$. Thus, the low-$T$ behavior can be modeled using the extended orbital scenario. 

For a quasi-2D Dirac system, the extended orbital contribution can be expressed as \cite{maebashi2019nuclear}
\begin{multline} \label{extended}
\bigg(\frac{1}{T_1T}\bigg)_{\perp c}=\frac{3}{2}\bigg(\frac{1}{T_1T}\bigg)_{\parallel c}=\frac{\mu_0^2\gamma_n^2e^2k_B}{(4\pi)^2} \\
\times \int\limits_{|E|>\Delta} dE\bigg[-\frac{\partial f(\varepsilon)}{\partial \varepsilon}\bigg]\frac{\sqrt{\varepsilon^2-\Delta^2}}{\hbar^2cv_F}\ln\frac{2(\varepsilon^2-\Delta^2)}{\hbar\omega_0|\varepsilon|},
\end{multline}
with $\varepsilon=\pm\sqrt{{v_F}^2k^2+\Delta^2}$ and $c$ the distance between nearest neighbor layers. In addition, $f(\varepsilon)$ is the Fermi function and $E_g=2\Delta$ is the gap. In the low-$T$ limit assuming $\Delta$ is small, this readily evaluates to $\frac{(\mu_0\gamma_n e)^2}{(4\pi)^2}\frac{k_B\mu}{\hbar^2cv_F}\ln(\frac{2\mu}{\hbar\omega_0})$. Comparing to the result \cite{maebashi2019nuclear} for a 3D point node in the same limits, $\frac{8\pi}{3}\frac{(\mu_0\gamma_n e)^2}{(4\pi)^2}\frac{k_B\mu^2}{\hbar^3v_F^2}\ln(\frac{2\mu}{\hbar\omega_0})$, $1/T_1T$ for the quasi-2D case is the same as the 3D case multiplied by a factor $\frac{3}{8\pi}\frac{\hbar v_F}{\mu c}$. Taking $\mu = 10$ meV, $v_F = 0.67 \times 10^6$ m/s, and $c = 6.7$ \si{\angstrom} for ZrTe$_2$, this is a factor of 7, with the quasi-2D situation enhanced essentially because of the increased phase space for the scattering phenomena leading to Eq.~(\ref{extended}), which can include events with $\Delta k$ covering the entire Brillouin zone in the direction perpendicular to the layers. With the low-$T$ $(1/T_1 T)_{\perp c}$ larger by a factor of about 10 as compared to that of the comparable point-node material ZrTe$_5$ \cite{tian2019dirac}, this indeed makes it plausible that the extended-orbital mechanism for high-mobility Dirac electrons is the dominant mechanism at low temperatures. In the low-$T$ limit, the ratio $(1/T_1T)_{\perp c}/(1/T_1T)_{\parallel c}$ is smaller than the expected 3/2 given by this model; however, note that Eq.~(\ref{extended}) was derived in the low-field limit, and it seems possible that such effects might renormalize the $(1/T_1T)_{\parallel c}$ results. In addition, while the normal-electron pocket at M is strongly dominated by Zr $d$-orbitals, a nonzero contribution due to Te states might also lead to a slowly varying background contribution to $1/T_1T$.

As shown in Fig.~\ref{discussion}(b), we determined that Dirac spins can give a considerable contribution to Knight shift due to core polarization combined with Landau level splitting for $B\parallel c$. However, we expect the core polarization mechanism to give a rather negligible contribution to $1/T_1T$. This can be seen from the Korringa relation \cite{carter1977metallic} which can provide an approximate upper limit for the spin $1/T_1T$. For $^{125}$Te, the Korringa relation will be, $(1/T_1T)_\mathrm{spin} = K_\mathrm{spin}^2/[2.6\times10^{-6}$ (s\,K)$^{-1}$], and with $|K_\mathrm{spin}|$ at low temperatures determined to be somewhat less than 100 ppm, choosing 100 ppm yields a limiting value $(1/T_1T)_\mathrm{spin} = 4\times10^{-3}$ (s\,K)$^{-1}$. This is considerably smaller than what is observed. Note also that in the low-$T$ limit where the Dirac spins are heavily polarized, the probability of spin-flip scattering can be reduced, further limiting $1/T_1T$. However, the extended orbital $1/T_1T$ due to high-mobility electrons is not connected to the shift via a Korringa relation, and from these considerations we determine that the spin-lattice relaxation rate of ZrTe$_2$ is dominated by this orbital contribution. These results will extend across the whole temperature range.

As the temperature increases past 10 K, $(1/T_1T)_{\parallel c}$ drops rather suddenly, reaching a minimum at about 40 K. This also coincides with a reported drop in the Dirac-carrier mobility, before the high-$T$ regime sets in with different behavior \cite{wang2019magnetotransport}. We believe that the change in $(1/T_1T)_{\parallel c}$ can be understood in terms of carrier scattering effectively reducing the dimensionality of the relaxation mechanism. Ref.~\cite{knigavko2007divergence} shows that the orbital $1/T_1T$ process due to high-mobility electrons, which relies upon a logarithmic divergence in the hyperfine coupling mechanism at large distances, will begin to cut off at a distance corresponding to the mean free path ($\ell$) as the scattering rate increases, so that $1/T_1T$ becomes proportional to $\ln(\ell)$. With little or no dispersion for the nodal-line carriers in the direction perpendicular to the layers, the mean free path will certainly be highly anisotropic. Once this length becomes considerably reduced, $1/T_1T$ will go over to the 2D case, for which the extended orbital $(1/T_1T)_{\perp c}$ is unchanged but $(1/T_1T)_{\parallel c}$ in this mechanism vanishes \cite{dora2009unusual,maebashi2019nuclear}. This is not to say that the layers become completely decoupled; a large reduction in mean free path is sufficient for this change to occur.

Above the minimum, $(1/T_1T)_{\parallel c}$ again starts to increase. As seen in the inset of Fig.~\ref{discussion}(b), the increase vs $T$ is also accompanied by a drop in chemical potential to maintain charge balance given the large $g_\mathrm{CB}(\varepsilon)$ contribution. As shown in Figs.~\ref{dft}(a) and (b), there is a split-off band at $\Gamma$ just below the Dirac node, which is more strongly dominated by Te $p$-electrons. As $\mu$ decreases, holes will begin to appear in these states, with a significant effect on the $^{125}$Te NMR because of their orbital weight. Aside from the $1/T_1T$ changes, there is also a change of character for the $T$-dependence of $K$, with a small increase in shift appearing for $B\perp c$. This behavior matches the observed change in magnetotransport behavior at these temperatures \cite{wang2019magnetotransport}, which we believe is a Lifshitz transition corresponding to the chemical potential meeting this split-off band edge. To understand the increase in $(1/T_1T)_{\parallel c}$ at high temperatures, we show in the Appendix that in addition to the extended orbital contribution, there is local orbital contribution \cite{knigavko2007divergence} to $1/T_1T$, which does not rely on logarithmic divergence at extended distances which will be larger for the $B\parallel c$ orientation as long as the Te $p_z$ contribution exceeds the Te $p_x$ and $p_y$ contributions [Fig.~\ref{discussion}(c)], which seems to be the case here. \textcolor{black}{Therefore, the high-temperature behavior can be understood in terms of an enhanced local-orbital contribution of $1/T_1T$, dominated by the split-off band which comes into play at higher temperatures, while the extended orbital contribution decreases as a consequence of the large decrease in carrier mobility.}

\section{Conclusions}

In conclusion, the topological nature of transition metal dichalcogenide ZrTe$_2$ is revealed here as a quasi-2D Dirac semimetal with a nodal line between $\Gamma$ and A. For magnetic fields perpendicular to the ZrTe$_2$ layers, the measured shift can be well-modeled by a combination of orbital shift and spin shift due to high mobility Dirac carriers. We also show that the low-temperature behavior of the spin-lattice relaxation rate can be explained through a quasi-2D Dirac electron model. In the intermediate temperature range, an increase in scattering of the Dirac carriers is applied to interpret the observed fast drop of the spin-lattice relaxation rate for the $B\parallel c$ orientation. With temperature further increasing, the local orbital contribution starts to dominate the spin-lattice relaxation rate with the significant contribution of a split-off band.

\begin{acknowledgments}
This work was supported by the Robert A. Welch Foundation, Grant No. A-1526 and Texas A\&M University.
\end{acknowledgments}

\appendix*

\section{Spin-lattice relaxation due to orbital and dipolar interactions}

The local orbital contribution to $1/T_1T$ is the mechanism typically associated with orbital hyperfine coupling in normal metals. As opposed to the extended-orbital mechanism \cite{lee1991relaxation,knigavko2007divergence}, the local contribution is expected to be limited to orbitals belonging to the atom containing the nucleus being measured. Following the treatment of Obata \cite{obata1963nuclear}, here we extend the calculation of $1/T_1T$ to $p$-electrons in the tetragonal symmetry corresponding to the 3-fold uniaxial symmetry for Te sites in ZrTe$_2$.

In the tight-binding approximation, the Bloch eigenfunctions are built up from localized atomic functions. For $p$-electrons, there are three independent orbital functions $p_x$, $p_y$ and $p_z$. With magnetic field $B$ along a certain direction, in our case $x$ and $z$, here are the mixed wavefunctions for uniaxial symmetry (omitting the product spin states):
\begin{equation} \label{mixwave}
 \Psi=
   \begin{dcases}
     \alpha^{1/2}p_z+(1-\alpha)^{1/2}\frac{1}{\sqrt{2}}(p_x+p_y), & B\parallel c\\
     \alpha^{1/2}p_y+(1-\alpha)^{1/2}\frac{1}{\sqrt{2}}(p_z+p_x), & B\perp c
   \end{dcases}
\end{equation}
where $\alpha$ is a parameter specifying the relative amount of $E$ symmetry ($p_x$ and $p_y$) vs $A_1$ symmetry ($p_z$) for magnetic field along $z$ (similarly for $B\perp c$ with $\Psi$ rotating correspondingly). For $B\parallel c$, when $\alpha=0$, the wavefunction contains only $p_x$ and $p_y$. With $\alpha=1$, only $p_z$ remains. For both dipolar interaction and orbital interaction contributions, we can thus determine the expressions of the corresponding spin-lattice relaxation rates, starting with a golden-rule relation, for which $1/T_1=2W=4\pi/\hbar k_BT \langle|\Psi |\mathcal{H}| \Psi \rangle|^2 g^2 (\varepsilon_F)$, where $\mathcal{H}$ is the orbital or dipolar hyperfine interaction Hamiltonian \cite{obata1963nuclear}, both of which are proportional to $1/r^3$ allowing the relative magnitudes to be readily compared. Also $g(\varepsilon_F)$ denotes the partial density of states at $\varepsilon_F$ for the Te $p$-orbitals, which are assumed to appear in the relevant band according to the amplitudes given in Eq.~(\ref{mixwave}). We obtain the following for the case for dipolar interaction:
\begin{equation} \label{dip}
\begin{split}
W_\mathrm{dip} 
 =&\frac{4\pi}{5}C\bigg(\bigg| \int_0^{2\pi}\int_0^{\pi} \Psi \Psi^*\frac{1}{2}Y_2^0 \sin\theta d\theta d\phi \bigg|^2 \\
& + \bigg| \int_0^{2\pi} \int_0^{\pi} \Psi \Psi^*\frac{\sqrt{3}}{2}Y_2^{-1} \sin\theta d\theta d\phi  \bigg|^2 \\
& + \bigg| \int_0^{2\pi} \int_0^{\pi} \Psi \Psi^*\sqrt{\frac{3}{2}}Y_2^{-2} \sin\theta d\theta d\phi  \bigg|^2 \bigg) \\
  = &
   \begin{dcases}
     \frac{C}{50}(9\alpha^2-12\alpha+5) & (B\parallel c)\\
     \frac{C}{200}(9\alpha^2+6\alpha+5) & (B\perp c),
   \end{dcases}
\end{split}
\end{equation}
where $\Psi$ is the wavefunction from Eq.~(\ref{mixwave}). Here $C=2\pi(\gamma_e\gamma_n\hbar^{3/2})^2g^2(\varepsilon_F)k_BT\langle r^{-3}\rangle^2$, where $\langle r^{-3}\rangle$ comes from the radial parts of the integrations which are not displayed in Eq.~(\ref{dip}). The integrals can be analytically evaluated giving the results also shown in Eq.~(\ref{dip}). For the case of the orbital interaction, the corresponding relations are
\begin{equation}
\begin{split}
  W_\mathrm{orb}
  = & \frac{C}{2}|\langle \Psi|l^{-1}|\Psi\rangle|^2 \\
  = & 
   \begin{dcases}
     2C\alpha(1-\alpha) & (B\parallel c)\\
     \frac{C}{2}(1-\alpha^2) & (B\perp c).
   \end{dcases}
\end{split}
\end{equation}
These results are shown in Fig.~\ref{discussion}(c) in the main text. As anticipated \cite{obata1963nuclear} the orbital term dominates in almost all cases. Also there is a crossing of terms at $\alpha=1/3$ which represents an equal mixture of orbitals, as expected since such a mixture becomes isotropic. When $\alpha$ is larger than $1/3$, the local orbital contribution for $B\parallel c$ exceeds that for $B\perp c$.

\end{document}